%% file: main2.tex
\documentclass[pdflatex,sn-mathphys-num]{sn-jnl}

\usepackage{graphicx}%
\usepackage{multirow}%
\usepackage{amsmath,amssymb,amsfonts}%
\usepackage{amsthm}%
\usepackage{mathrsfs}%
\usepackage[title]{appendix}%
\usepackage{xcolor}%
\usepackage{textcomp}%
\usepackage{manyfoot}%
\usepackage{booktabs}%
\usepackage{algorithm}%
\usepackage{algorithmicx}%
\usepackage{algpseudocode}%
\usepackage{listings}%
\usepackage{booktabs}%
\usepackage{verbatim}%

\theoremstyle{thmstyleone}%
\theoremstyle{thmstyletwo}%

\theoremstyle{thmstylethree}%

\begin{document}
\include{authors.tex}

\title[Measurement of 80-200~MeV/n $^{16}$O nuclear cross-section on Carbon and Polyethylene targets with the nuclear emulsion detector of the FOOT experiment]{Measurement of 80-200~MeV/n $^{16}$O nuclear cross-section on Carbon and Polyethylene targets with the nuclear emulsion detector of the FOOT experiment}






\abstract{Accurate knowledge of nuclear fragmentation cross-sections is essential for optimizing charged particle therapy. In this study, conducted within the framework of the FOOT (FragmentatiOn Of Target) experiment, we present the first measurements with a large angular acceptance of total charge-changing cross-section and the cross-section for the production of fragments (production cross-section) for $^{16}$O ions interacting with Carbon (C) and Polyethylene (C$_2$H$_4$) targets in the kinetic energy range of 80 to 200 MeV/nucleon. Measurements were performed using the Emulsion Cloud Chamber (ECC) technique, which combines high spatial resolution and angular acceptance, up to 45$^\circ$. The results are compared with Monte Carlo model predictions. Moreover, the total charge-changing and fragment production cross-sections for $^{16}$O on Hydrogen in the same energy range are derived.}




\maketitle

\section{Introduction}\label{sec1}

The knowledge of nuclear reaction cross-sections is crucial for an accurate evaluation of the dose in both Charged Particle Therapy 
(CPT)~\cite{Durante2016} and space radiation protection (SRP)~\cite{RevModPhys.83.1245}. Numerous studies have investigated the production of particles resulting from the fragmentation of primary ions; however, experimental data remain scarce in the kinetic energy range of 100--400 MeV/n, which is relevant for medical applications, and above 700 MeV/n, which is important for SRP\footnote{An up-to-date database with all published data concerning total nuclear reaction cross-sections together with comparisons between experimental data and transport codes can be found in~\cite{luoni2021total}.}.

To address this gap, the FOOT (FragmentatiOn Of Target) experiment~\citep{battistoni2021} was designed to measure the target and projectile fragmentation induced by protons and light ions on nuclei commonly found in human tissues. 

In this paper, we report the first FOOT measurement of $^{16}$O nuclear fragmentation cross-sections at energies ranging from 80 MeV/n to 200 MeV/n, using a setup capable of detecting fragments emitted at polar angles up to 45$^\circ$. 
This large angular acceptance enables a more complete characterization of the fragmentation process, improving the accuracy of cross-section measurements.




\section{Material and methods}\label{sec2}
The FOOT experiment measures projectile and target fragmentation in ion collisions with the most abundant elements in biological tissues. Of particular interest are the proton-nucleus (p-N) collisions, which are challenging to detect because of the extremely short range of the resulting fragments, of the order of a few micrometers. Target fragmentation cross-sections are hence measured using an inverse kinematic approach, studying the interactions of a heavy ion beam impinging on pure and hydrogen-rich targets~\cite{battistoni2021}. 

The 
physics program of the experiment focuses on the nuclear fragmentation of $^4$He, $^{12}$C and $^{16}$O ions on Carbon (C) and Polyethylene (C$_2$H$_4$) targets, in the energy range 100-800 MeV/n.

Two complementary setups are employed: a magnetic spectrometer focusing on $Z\ge2$ forward emitted fragments ($\le 8^{\circ}$) and an emulsion spectrometer for $Z\le3$ fragments over a wide angular range ($\le 45^{\circ}$). The overall detector design, expected performances 
and results from preliminary test beams are described in recent works~\citep{battistoni2021, Dong_1}. Cross-sections for a 400~MeV/n $^{16}$O beam on a C target have already been published by FOOT~\cite{nmw9-ldrm,10.3389/fphy.2022.979229} using a setup dedicated to forward fragment production for angles up to 6$^\circ$. 

In this work we present the results obtained with the nuclear emulsion spectrometer. 

\subsection{The nuclear emulsion spectrometer}\label{subsec2}

Nuclear emulsions are special photographic films composed of 
AgBr silver halide crystals dispersed in an organic gelatine. 
When a charged particle crosses the sensitive layer, the crystals along its trajectory are sensitized and, after chemical development, form a sequence of silver grains. The grain density of a track stores information on the locally deposited energy. 
The films used, produced by Nagoya University and Slavich Company, feature 70~\textmu m sensitive layers on a 210~\textmu m plastic base, with a sensitivity of 30~grains/100~\textmu m for Minimum Ionizing Particles (MIP). 
After development, the analog images are digitized by the New Generation Scanning System (NGSS) optical automatic microscopes~\citep{alexandrov2015, alexandrov2016, alexandrov2017}.



Nuclear emulsion spectrometers used by FOOT are based on Emulsion Cloud Chambers (ECC)~\citep{PhysRev.85.900}, which consist of interleaved nuclear emulsion films and passive materials, forming a compact sampling calorimeter for high-resolution charged track reconstruction.

The characteristics of the ECCs used for the FOOT experiment have been detailed in~\cite{montesi2019, Galati2021, galati2024charge}, and are briefly summarized here for clarity.

\begin{figure}[h]
\centering
\includegraphics[width=0.8\textwidth]{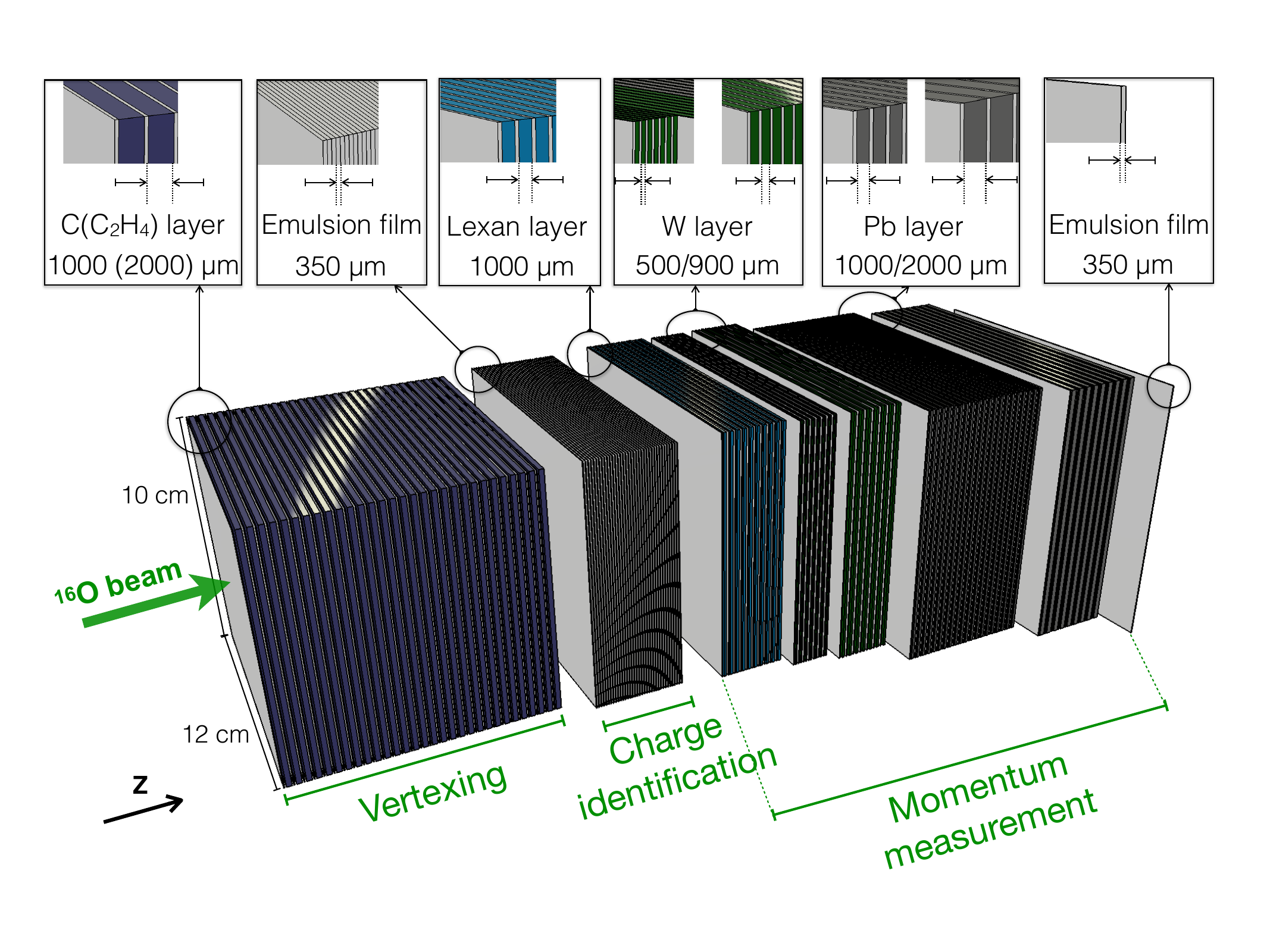}
\caption{The spectrometer features three sections optimized for vertex reconstruction, fragment charge identification, and momentum measurement. Emulsion films and passive layers are housed in a light-shielded, 3D-printed box ($12~\text{cm} \times 14~\text{cm} \times 30~\text{cm}$) featuring a $5.7~\text{cm} \times 5.7~\text{cm}$ beam-entry window.}\label{fig:ECC1.pdf}
\end{figure}

Each ECC is composed of three sections (see Figure~\ref{fig:ECC1.pdf}). The first one (S1) is designed to reconstruct vertices and consists of 30 alternating layers of nuclear emulsion films and target material: natural Carbon (1~mm thick, $\rho = 1.73$~g/cm$^{3}$) or Polyethylene C$_2$H$_4$(2~mm thick, $\rho = 0.94$~g/cm$^{3}$). 
Hereafter, we refer to ECC1 as the detector with the Carbon target and to ECC2 as the one with the Polyethylene target. 
The second section (S2) is aimed at fragments' charge identification and 
the third one (S3) is dedicated to momentum measurement. 

\subsection{Experimental set-up at GSI}\label{subsec2}
Two exposures were performed in cave~A of the GSI Helmholtzzentrum f\"ur Schwerionenforschung~facility in Darmstadt (Germany), using a $^{16}$O beam with kinetic energy of 200~MeV/n~\citep{nmw9-ldrm, ubaldi2025measurements}. The beam particles were monitored by the Start Counter, a detector based on a thin plastic scintillator~\citep{Traini2020Cimento, battistoni2021} and by the Beam Monitor, which consists of a drift chamber~\citep{DONG2021164756}. 
Ions entered perpendicularly to the emulsion films, defining the reference axis for fragment emission. To balance statistics and pile-up avoidance, a Gaussian $^{16}$O beam (6 mm FWHM) was scanned over a $24\times24$ mm$^2$ area using a $25\times25$ spot squared spiral grid (1 mm steps). The nominal fluence was 3100~ions/cm$^2$ and 3200~ions/cm$^2$ for C and C$_2$H$_4$ targets, respectively.

\subsection{Tracks and vertices reconstruction}\label{reco}
\begin{figure}
	\centering 
	\includegraphics[width=0.8\textwidth, angle=0]{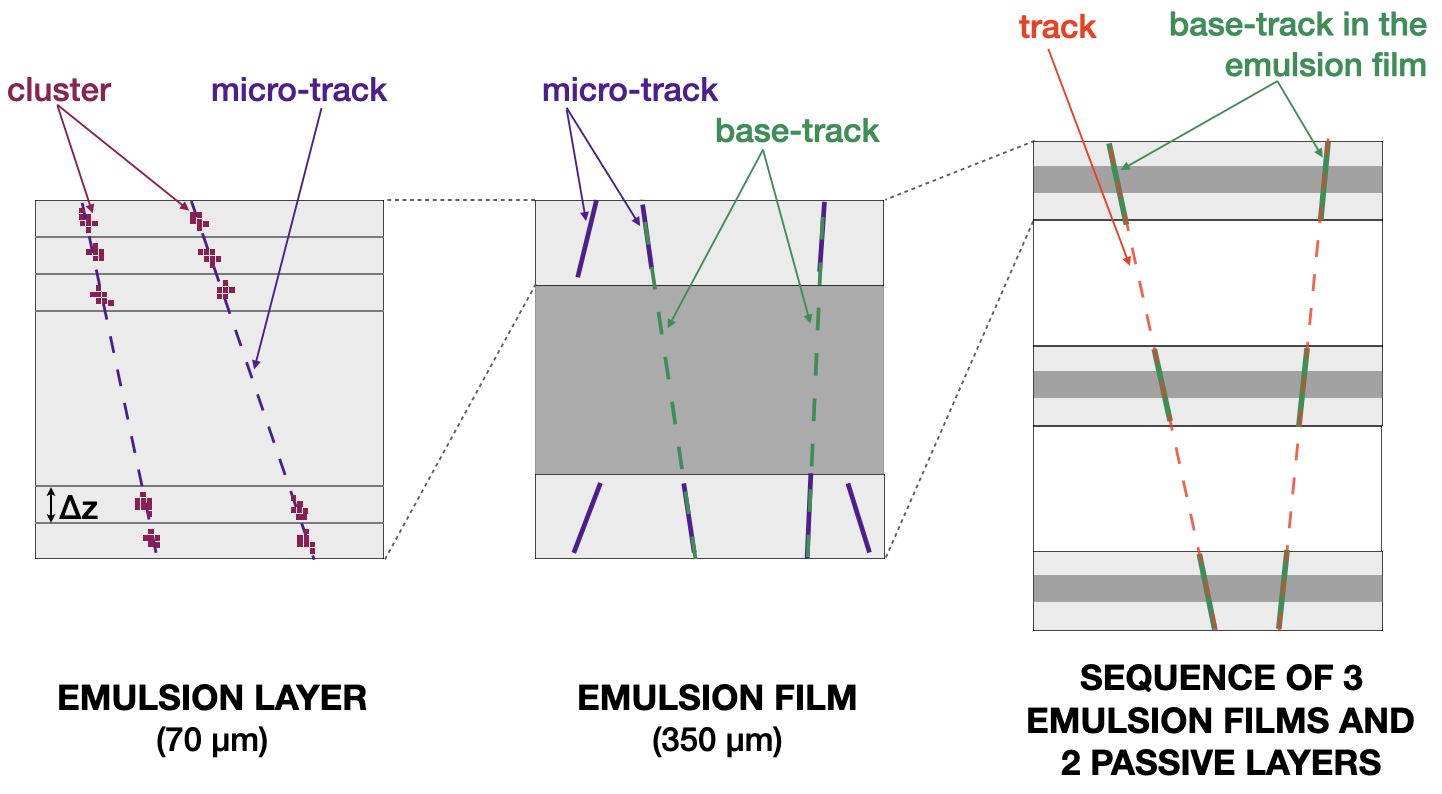}	
	\caption{Left: micro-track reconstruction from tomographic images at uniform depths ($\Delta z$) within a sensitive layer. Center: base-track formation by linking micro-tracks across adjacent sensitive layers. Right: full track construction via base-track association.\label{fig:emulsion}} 
\end{figure}

The analysis software is based on the FEDRA framework~\citep{SySal2000,SySal.NET, LASSO, FEDRA}. Track reconstruction begins by identifying sequences of aligned grains in each emulsion sensitive layer, forming a \textit{micro-track}. 
Subsequently, micro-tracks from the top and bottom layers of each emulsion film are linked to form \textit{base-tracks}~\cite{Arrabito:2007td}.
Before track reconstruction, an alignment routine determines the optimal affine transformation to ensure the best match between base-tracks from adjacent films. 
A Kalman-filter tracking algorithm connects base-tracks into \textit{volume-tracks} (simply ``tracks"). A schematic sketch of the track reconstruction procedure is shown in fig.~\ref{fig:emulsion} and a more detailed explanation can be found in ref.~\cite{Galati2021}. Base-tracks with $\tan\theta<1$ were recorded. The polar angle was measured with respect to the \textit{z} axis, perpendicular to the surface of the emulsion films. 
During tracking, finer alignment is achieved using long tracks, resulting in the creation of a global reference system. 

The detection efficiency of a nuclear emulsion film is evaluated by counting how many base-tracks were not detected in passing tracks. The average detection efficiency $\phi_{det}$ measured in S1 of each detector is $80\%$ for ECC1 and $90\%$ for ECC2. It should be noted the value of $\phi_{det}$ summarizes several processes related to the emulsion films' chemical development and scanning, as well as residual inaccuracies in the alignment and tracking process.  



FEDRA's vertex finding algorithm~\cite{FEDRA} was employed to find tracks that originated from interactions of the primary ions.

The vertexing procedure is based on track clustering according to a cut on the impact parameter, i.e.~the minimum distance between the extrapolated track and the vertex. Converging tracks starting inside the ECC and having an impact parameter below 30~\textmu m were recognized as fragments coming from the interaction of a primary ion.

\subsection{Monte Carlo simulation of the experiment}\label{mc}

A Monte Carlo (MC) simulation of the experimental setup was performed using the FLUKA software~\citep{osti_877507,Ballarini2024}. The simulation serves two main purposes: comparison between experimental data and nuclear model predictions and evaluation of the reliability of the reconstruction software.

In the energy range of primary ions relevant to this work, nucleus--nucleus collisions are handled in FLUKA by two different models. For energies below about 150 MeV/n, the Boltzmann Master Equation (BME) model~\cite{cavinato98} is employed. In this approach, the time evolution of the many-body phase-space distribution of particles in nuclear matter is governed by a stochastic solution of the Boltzmann transport equation. The reaction dynamics is described as a sequence of elementary two-body processes, including elastic and inelastic scattering, excitation and decay of baryonic resonances, and secondary particle production. The FLUKA implementation~\cite{cerutti2006} samples from the results of the numerical integration of the BME. 
At projectile energies exceeding about 150 MeV/n, a gradual transition is made to an interface with a modified rQMD-2.4 (relativistic Quantum Molecular Dynamics) model~\cite{Sorge1,Sorge2}, which can also operate as an intra-nuclear cascade. Examples of FLUKA results obtained using the modified rQMD-2.4 model and compared with experimental data are reported in Refs.~\cite{andersen2004fluka,aiginger2005fluka,ballarini2007fluka}. Although originally developed for higher energies, this model can be extended to cover the upper part of the energy range relevant for the FOOT experiment.  
Both BME and rQMD are coupled to a pre-equilibrium stage, which in FLUKA is governed by the PEANUT (PreEquilibrium Approach to Nuclear Thermalization) model~\cite{ferrari1997physics,ferrari2002nuclear}. This pre-equilibrium treatment is based on the Geometry-Dependent Hybrid (GDH) exciton model~\cite{blann1971hybrid}. The final de-excitation of the equilibrated nuclei following the nucleus--nucleus collision is handled by the FLUKA evaporation/fission/fragmentation module. For nuclei with $A<17$, the late stages of the interaction are modeled using a phase-space Fermi break-up model~\cite{fermi1950high,epherre1967calcul}.

The simulation geometry incorporated detailed models of the monitoring detectors (Start Counter and Beam Monitor~\cite{Dong_1}) as well as the emulsion spectrometer. In order to test the performance of the reconstruction algorithm, FLUKA MC hits were converted to the emulsion base track format. This conversion took into account several data-driven effects such as:
\begin{itemize}

    \item the finite spatial and angular resolutions of the detector, which were simulated by applying a Gaussian smearing of about 2~\textmu m to the true coordinates and 2.5~mrad to angles; 
    \item the average detection efficiency $\phi_{det}$, which was taken into account by randomly omitting~$20\%$ (ECC1) and $10\%$ (ECC2) base-tracks in each film; 
    \item  cosmic rays background, which was incorporated by adding detected base-tracks generated by cosmic rays in each film. These base-tracks were modelled  from those located in the area of each nuclear emulsion film outside the interaction region;
   
    \item affine transformations, which were derived from data and applied to each emulsion film to realistically simulate the misalignment between the films. 
    
\end{itemize}
Once the conversion was completed, alignment, tracking and vertex reconstruction were carried out using with the same reconstruction algorithms as for data.
The analysis selection criteria applied to tracks and interactions are described below:

\begin{itemize}
    \item charged particles must have a minimum momentum of 0.1~GeV/c and a maximum $\tan\theta$ of 1, i.e.~a maximum polar angle of 45$^\circ$ (visible tracks);
    \item at least two visible tracks (fragments) must originate from a vertex;
    
    \item at least two tracks associated with a vertex must have at least 3~base-tracks;
    
    \item at least one fragment associated with a vertex must reach the second section of the detector.
\end{itemize}
These selection criteria were applied both to data and to the MC sample. 
The fraction of reconstructed tracks corresponding to genuine particle trajectories, as opposed to spurious combinations of background noise, has been estimated from MC and it is higher than $99\%$, indicating a highly reliable track reconstruction. 
The fraction of tracks correctly associated with their originating vertex is approximately $90\%$, for both ECCs. 

Ten ECC detectors were simulated for each target, with ${2\times10^4}$ beam particles impinging on each one, resulting in a total statistics of ${2\times10^5}$ beam particles. This was done to accurately reproduce the track density conditions observed in the experimental data.

\begin{figure}
	\includegraphics[width=0.54\textwidth]{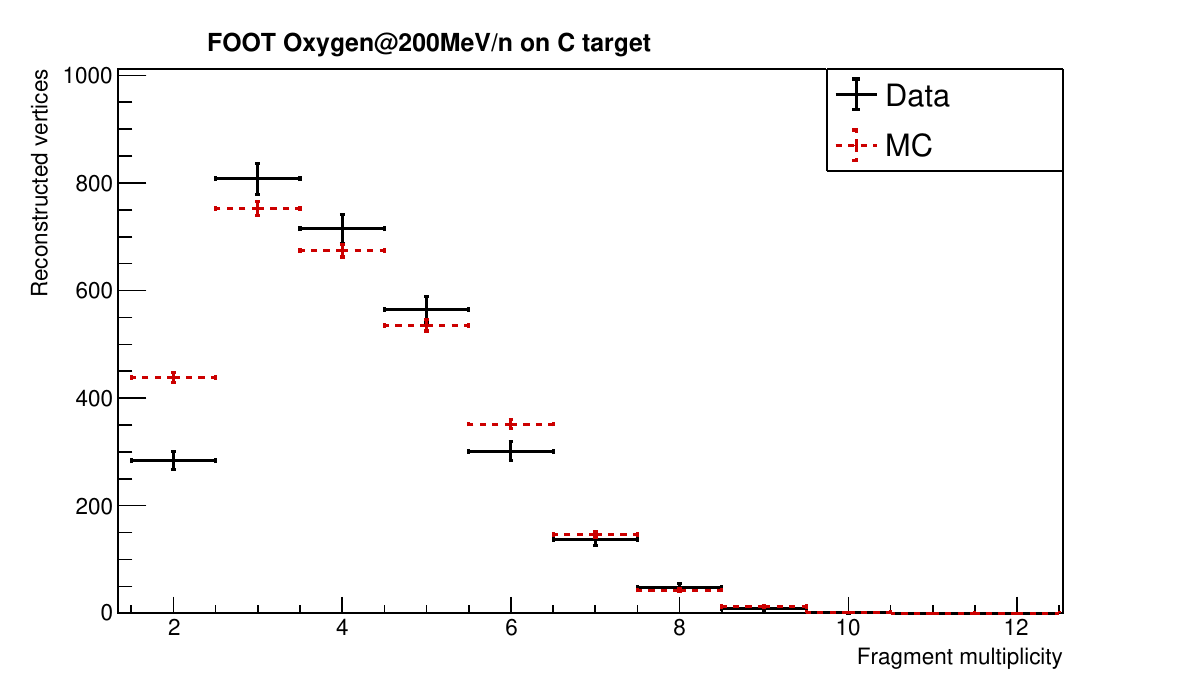}	
    \includegraphics[width=0.54\textwidth]{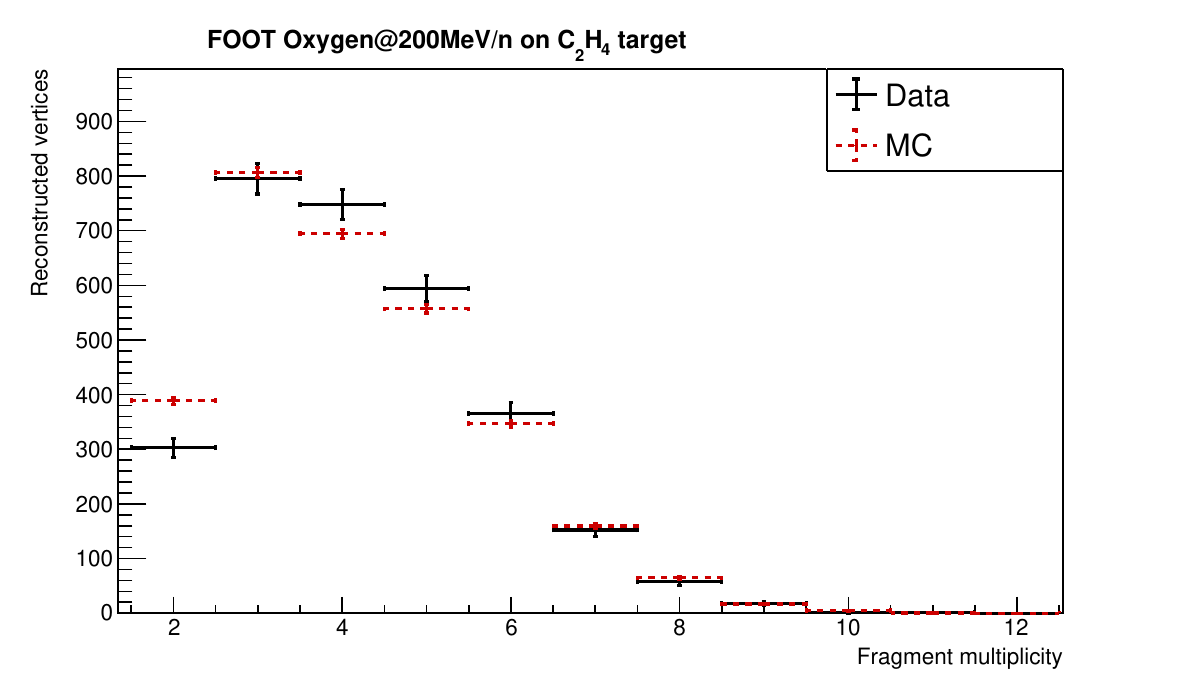}	
	\caption{Distribution of fragment multiplicity for ECC1 (left) and ECC2 (right).}
	\label{fig:mult}
\end{figure}

\begin{figure}
	\includegraphics[width=0.54\textwidth]{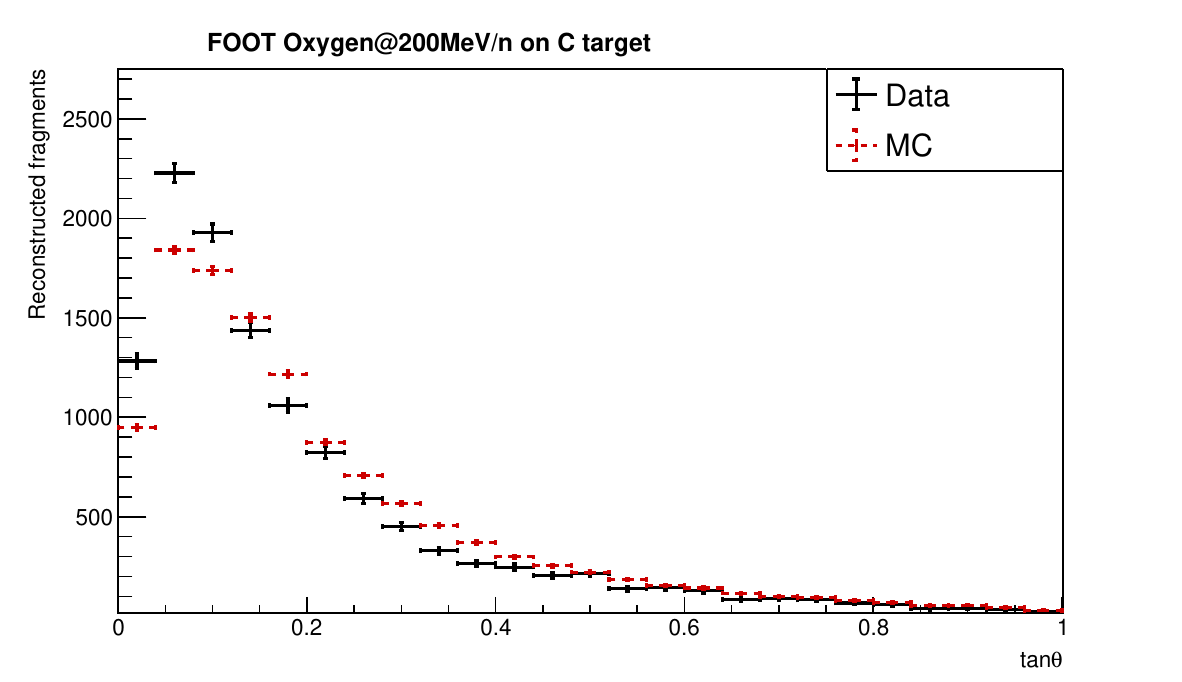}	
    \includegraphics[width=0.54\textwidth]{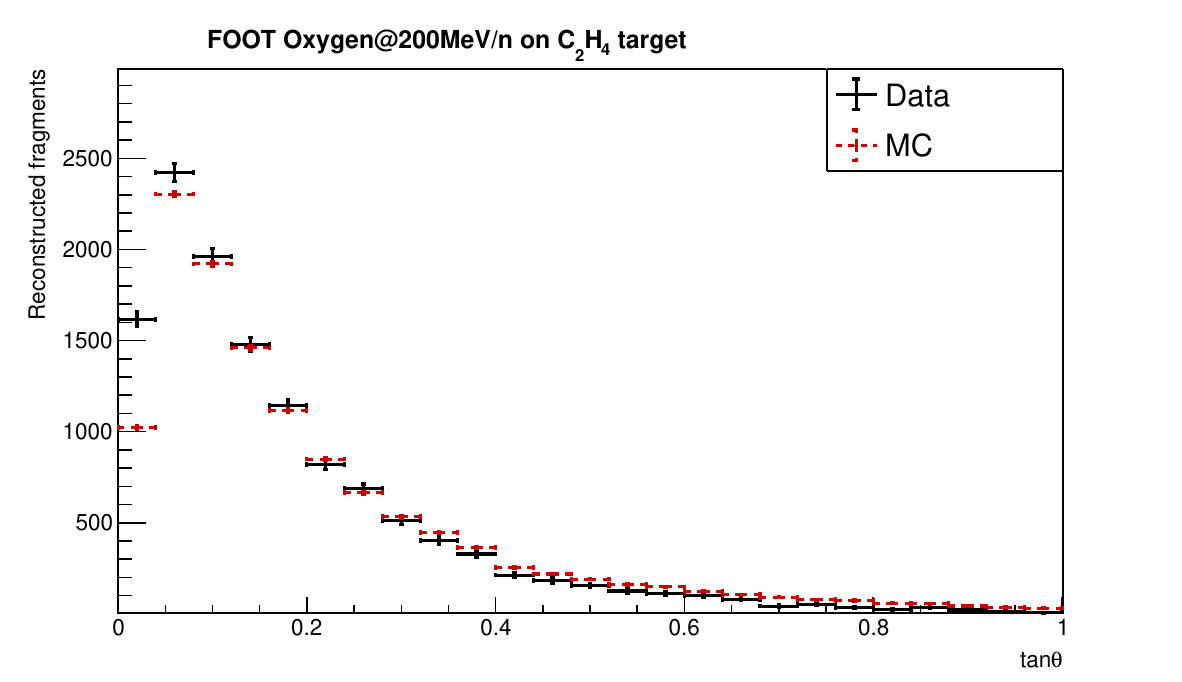}	
	\caption{Angular distribution of fragments associated with reconstructed vertices in ECC1 (left) and ECC2 (right).} 
	\label{fig:slope}
\end{figure}

The multiplicity of fragments originating from the reconstructed vertices in data and MC is reported in Fig.~\ref{fig:mult}, while the angular distributions of all fragments are shown in Fig.~\ref{fig:slope}.  
For direct comparison between the simulated and experimental results, all distributions have been scaled to match the total number of impinging particles.


\subsection{Cross-section evaluation}\label{subsec2}

Inelastic interactions between the projectile and target nuclei can either induce excitation processes within the nuclei or lead to their fragmentation, producing secondary nuclei from both the projectile and target. 
The total reaction cross-section measures the likelihood of any nuclear process occurring, while the fragment production cross-section specifically quantifies the probability of producing a fragment as a result of nuclear fragmentation~\cite{luoni2021total}. 
The measured charge-changing cross-section does not include reaction channels leaving the projectile charge unchanged, such as pure neutron-removal processes. 

\begin{figure}
	\centering 
\includegraphics[width=0.8\textwidth]{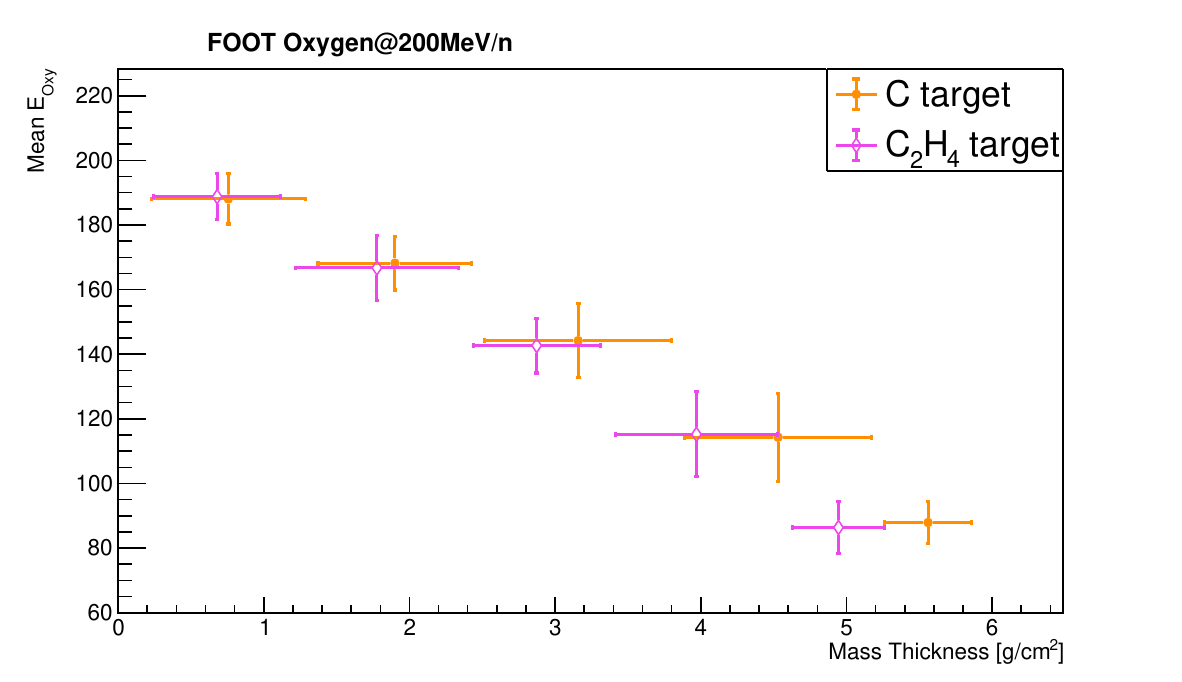}	
	\caption{Energy of the Oxygen ions as a function of the mass thickness in ECC1 and ECC2.} 
	\label{fig:oxy_ekin}
\end{figure}

Within each energy interval, the total charge-changing cross-section is determined from the number of reconstructed vertices, while the fragment production cross-section is based on the total number of charged nuclear fragments detected, summed over all reconstructed vertices. Both quantities are calculated using the following expression:

\begin{equation}\label{eq:xsec}
    \sigma(x) _{C\ or\ C_2 H_4} = \frac{Y(x)}{N_B(x) N_{TG} \epsilon_{\text{reco}}(x)}
\end{equation} 

where $x$ denotes the sub-section. In Eq.~\ref{eq:xsec}, the yield $Y(x)$ corresponds to the number of reconstructed interaction vertices for the total charge-changing cross-section, and to the total number of reconstructed charged fragments for the production cross-section.

The symbols and parameters used are reported in table~\ref{tab:symbols}.

\begin{table}[h]
\begin{tabular}{@{}ll@{}}
\toprule
\textbf{Symbol} & \textbf{Description} \\
\midrule
$x$ & sub-section of the detector in S1\\
$Y(x)$ & Number of vertices $Y_V$ (for total charge-changing cross-section) \\
 & or fragments $Y_F$ (for fragment production cross-section) in the $x$ sub-section\\
$N_B(x)$ & Number of ions colliding on the target in the $x$ sub-section\\
$N_{TG}$ & Number of target atoms/molecules per unit area: $\frac{\rho d N_A}{A}$, with: \\
$\rho$ & target density: \\
       & \quad $\rho_{C_{\text{nat}}}$ = 1.73 g/cm$^3$ \\
       & \quad $\rho_{C_2H_4}$ = 0.94 g/cm$^3$ \\
$d$ & target thickness: \\
    & \quad $d_{C_{\text{nat}}}$ = 0.1 cm per layer \\
    & \quad $d_{C_2H_4}$ = 0.2 cm per layer \\
$N_A$ & $6.022\cdot 10^{23}$/mol \\
$A$ & molar mass: \\
    & \quad $A_{C_{\text{nat}}}$ = 12 g/mol \\
    & \quad $A_{C_2H_4}$ = 28 g/mol \\
$\epsilon_{\text{reco}}(x)$ & reconstruction factor, obtained from the comparison between MC True and \\
& MC Reconstructed in the $x$ sub-section \\
\botrule
\end{tabular}
\caption{Parameters used in the cross-section evaluation.}
\label{tab:symbols}
\end{table}

The proton-nucleus (p-N) cross-sections on Hydrogen ($\sigma_H$) are derived by analysing the interactions with C$_2$H$_4$ and C targets. Since C$_2$H$_4$ consists of Carbon and Hydrogen atoms, the total cross-section measured for Polyethylene ($\sigma_{C_2H_4}$) can be expressed as a combination of contributions from Carbon and Hydrogen nuclei: $\sigma_{C_2H_4} = 2 \sigma_{C_{\text{nat}}} + 4 \sigma_H$. 
Using this relation, it is possible to compute the cross-section on Hydrogen ($\sigma_H$) by subtracting the Carbon contribution ($\sigma_{C_{\text{nat}}}$) from the total cross-section measured on C$_2$H$_4$ ($\sigma_{C_2H_4}$):

\begin{equation}
\sigma(x)_{H} = \frac{1}{4}\left[\sigma(x)_{C_2 H_4} - 2\sigma(x)_{C_{\text{nat}}}\right]
\end{equation}

The feasibility of the subtraction method was previously demonstrated in~\cite{GANIL}. 
An in-depth description of the methodologies employed in measuring the aforementioned quantities is provided in the following paragraphs.

\section{Data analysis}\label{sec2}

\subsection{$Y$ measurement}\label{Y}

The selection criteria used to reject the background have been listed in paragraph~\ref{mc}.

A dedicated procedure was implemented to handle cases of primary Oxygen fragmentation with very small kink angles, where the primary ion and the heavy fragment may be reconstructed as a single track, such as the ${O\rightarrow N+p}$ topology. When an Oxygen track is found beyond the Bragg peak, the track is inspected for anomalous angular deviations and, if necessary, split into two tracks and a corresponding vertex is reconstructed, provided quality criteria are satisfied.

For the cross-section evaluation, only vertices occurring in the target material are selected, while those occurring in the nuclear emulsion films are discarded based on the vertex positions. 

The number of vertices that meet the selection criteria has also been counted in the region outside the beam window in the experimental data, indicating that spurious vertices account for less than $0.2\%$. 

\begin{figure}
	\includegraphics[width=0.54\textwidth]{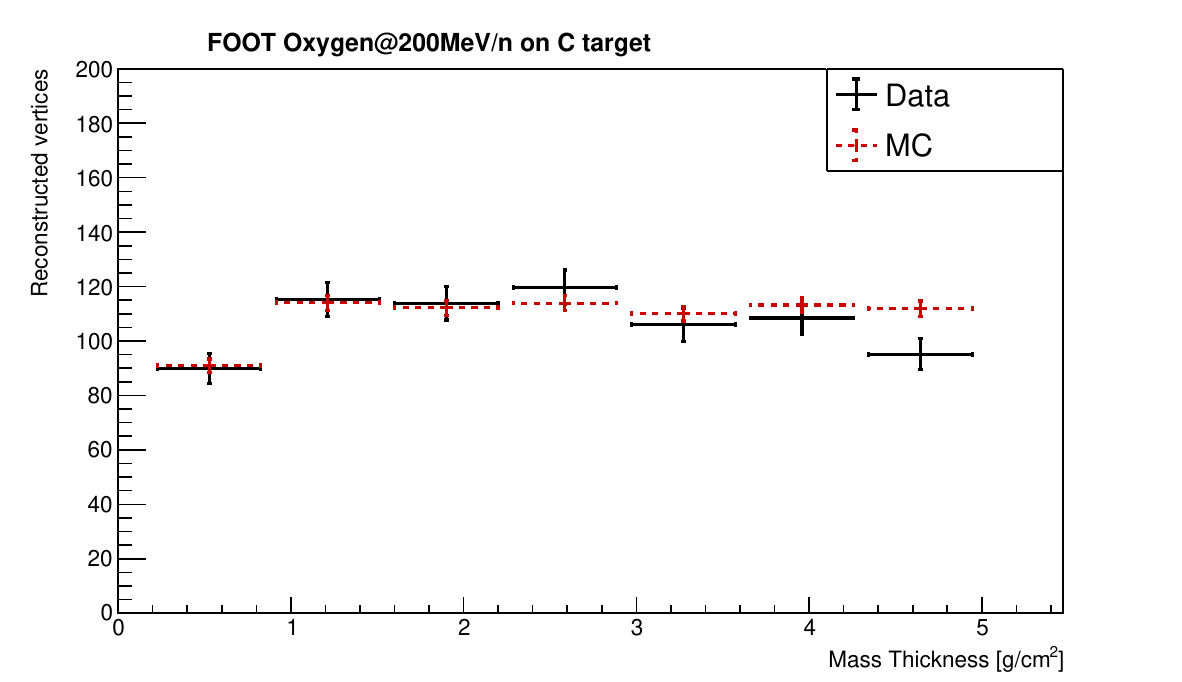}	
 \includegraphics[width=0.54\textwidth]{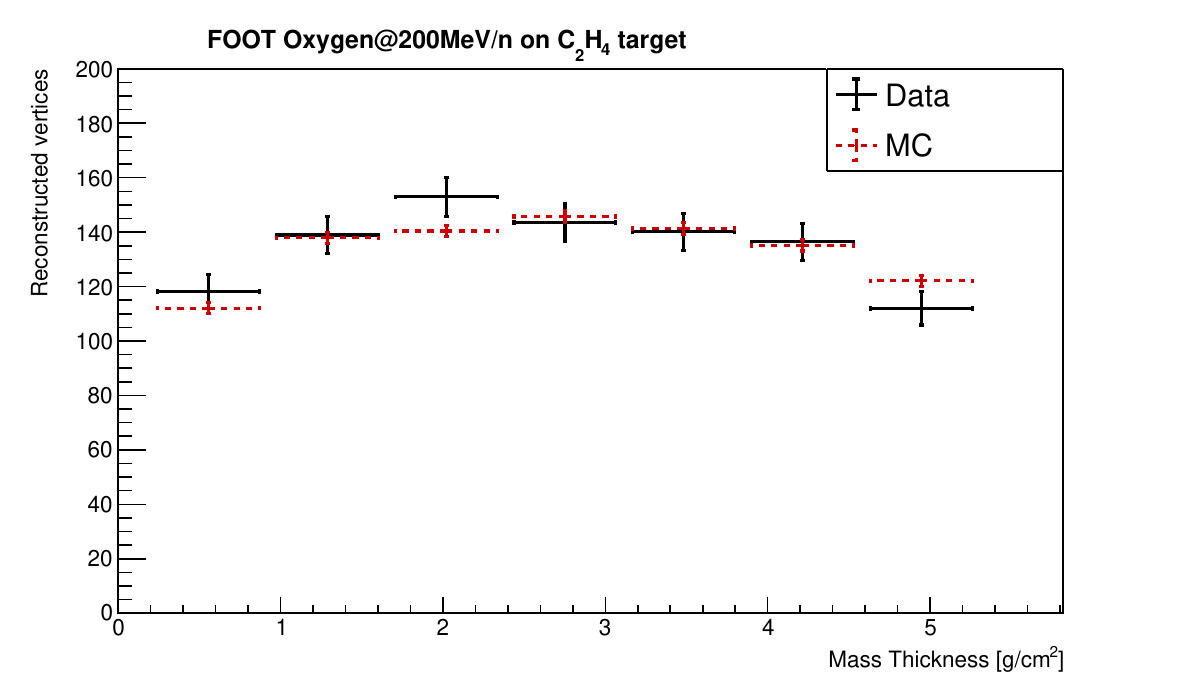}	
	\caption{Number of vertices $Y_V$ reconstructed in Section 1 of ECC1 (left) and ECC2 (right), as a function of the mass thickness.} 
	\label{fig:vtxs}
\end{figure}

\begin{figure}
	\includegraphics[width=0.54\textwidth]{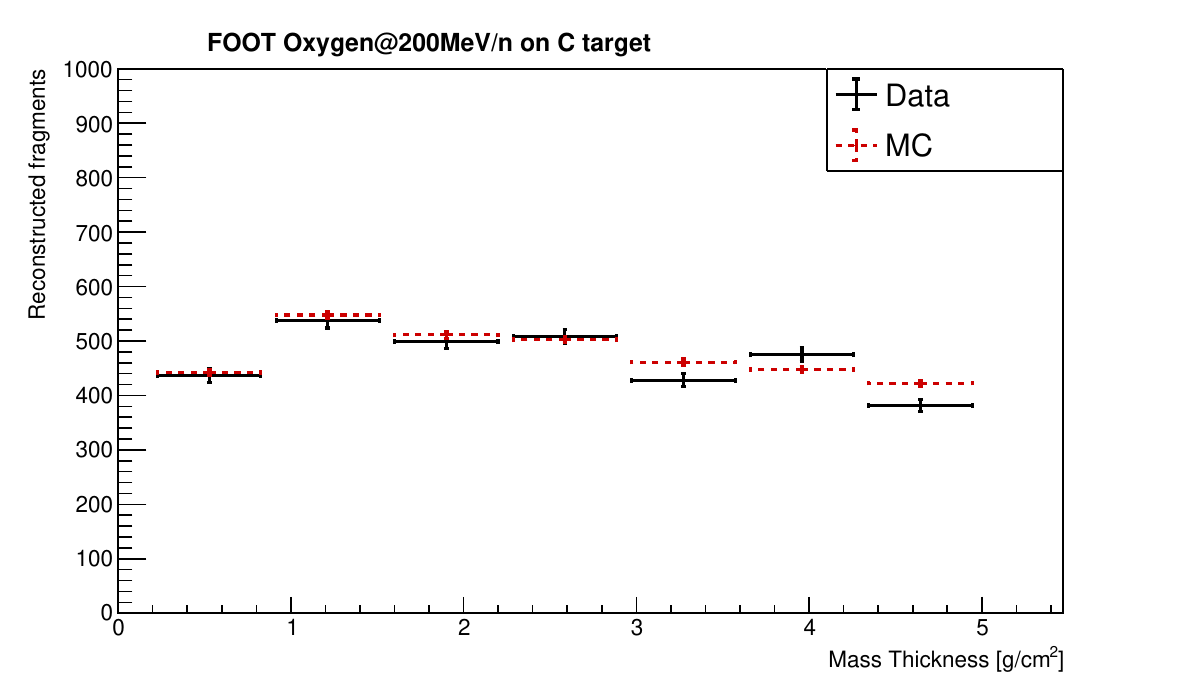}	
 \includegraphics[width=0.54\textwidth]{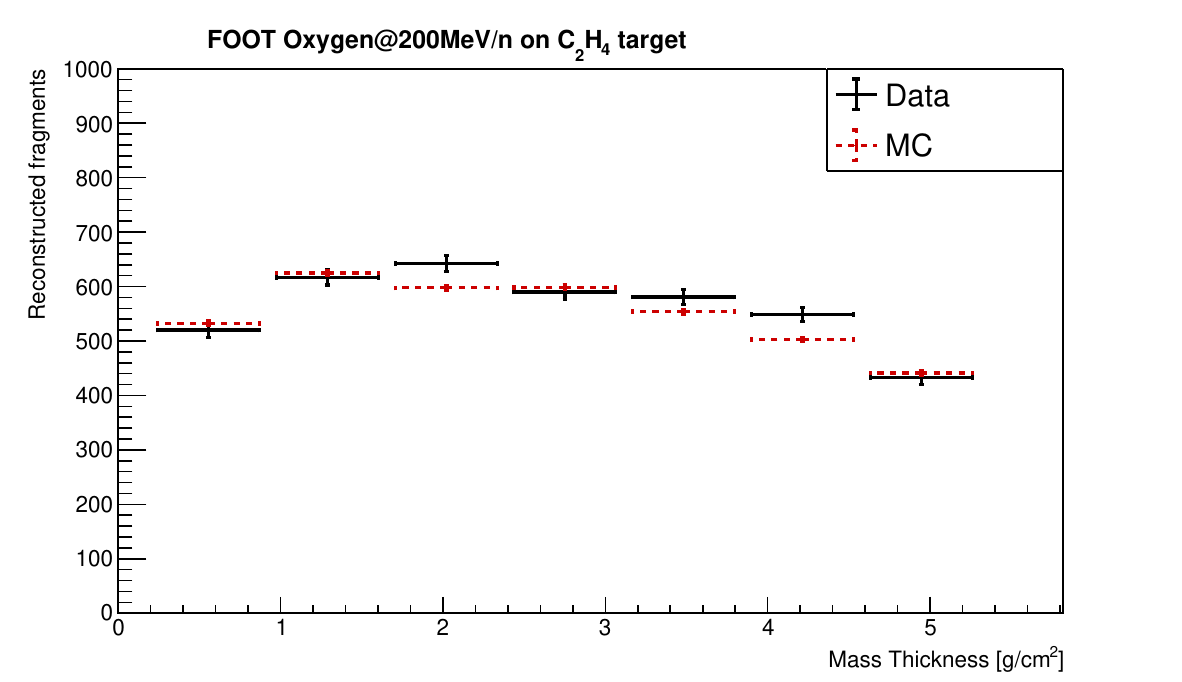}	
	\caption{Number of fragments $Y_F$ reconstructed in Section 1 of ECC1 (left) and ECC2 (right), as a function of the mass thickness.} 
	\label{fig:trks}
\end{figure}

The number of vertices $Y_V$ reconstructed in S1 is shown, for two different targets, in Figure~\ref{fig:vtxs} as a function of the integrated material traversed, defined as the product of the material thickness and its density. This parameter effectively quantifies the total amount of matter encountered by the particles. Similarly, Figure~\ref{fig:trks} displays the number of fragments $Y_F$ as a function of the same integrated material quantity.

\subsection{$N_B$ measurement}\label{NB}

\begin{figure}
	\centering
	\includegraphics[width=0.45\textwidth]{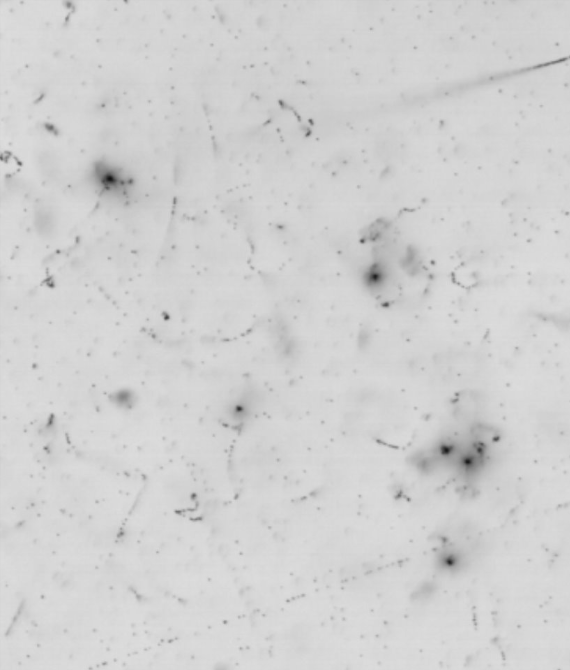}	
    \includegraphics[width=0.45\textwidth]{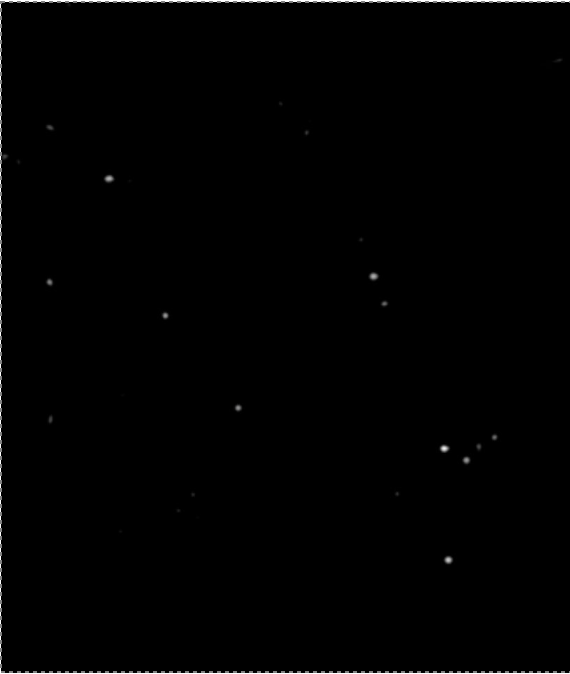}	
	\caption{
    Microscope images of Oxygen ion tracks. Left: raw image (${250 \times 275}~\mu\text{m}^2$) showing Oxygen ions as dark spots. Right: same image after applying the Gaussian filter.} 
	\label{fig:oxy_img}
\end{figure}

To determine the number of ions ($N_B$) colliding with each target, optical microscope images were processed using a high-pass Gaussian filter~\cite{aleksandrov2012methods}. The kernel size was optimized to match the typical dimensions of Oxygen-induced pixel clusters.  
This filtering technique enhances the detection efficiency of highly ionizing particles by effectively suppressing the signal from minimum ionizing particles, which produces smaller pixel clusters. Figure~\ref{fig:oxy_img} (left) shows a typical raw image of the microscope, where darker spots correspond to Oxygen ions entering the emulsion film perpendicularly. The same image, after the application of the high-pass Gaussian filter, is shown on the right of the Figure ~\ref{fig:oxy_img}, where high-ionization clusters are more clearly distinguishable. 
This processing enables reliable counting of the spots associated with highly ionizing particles. 

\begin{figure}
	\includegraphics[width=0.54\textwidth]{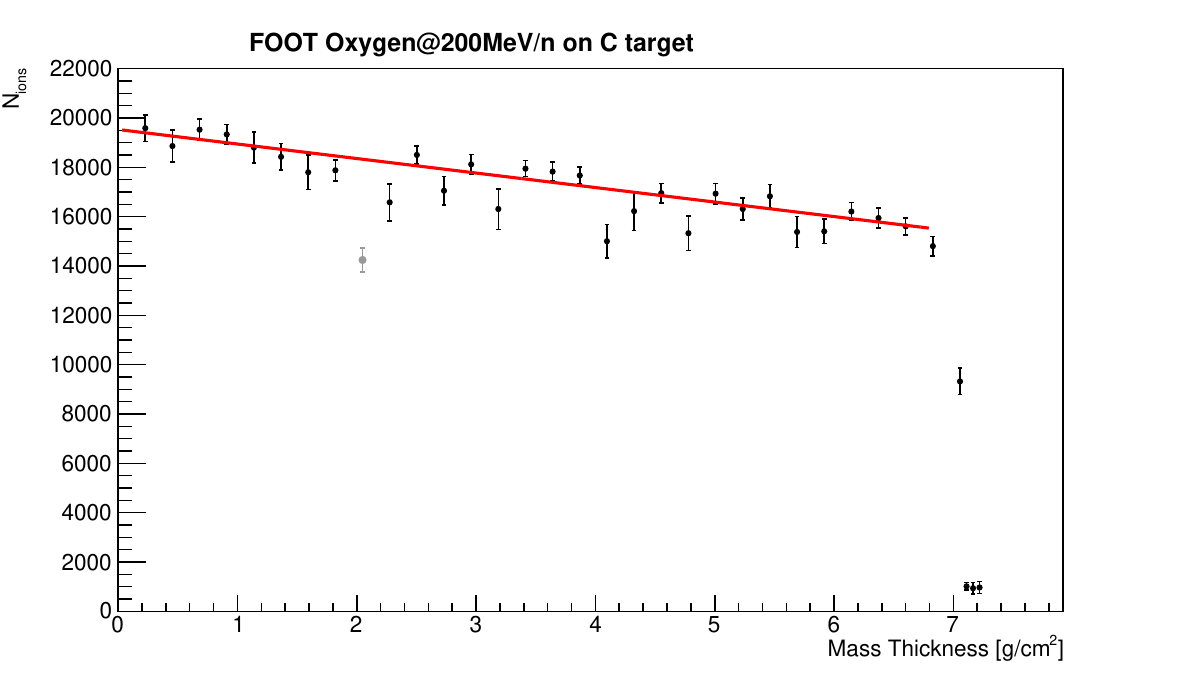}	
    \includegraphics[width=0.54\textwidth]{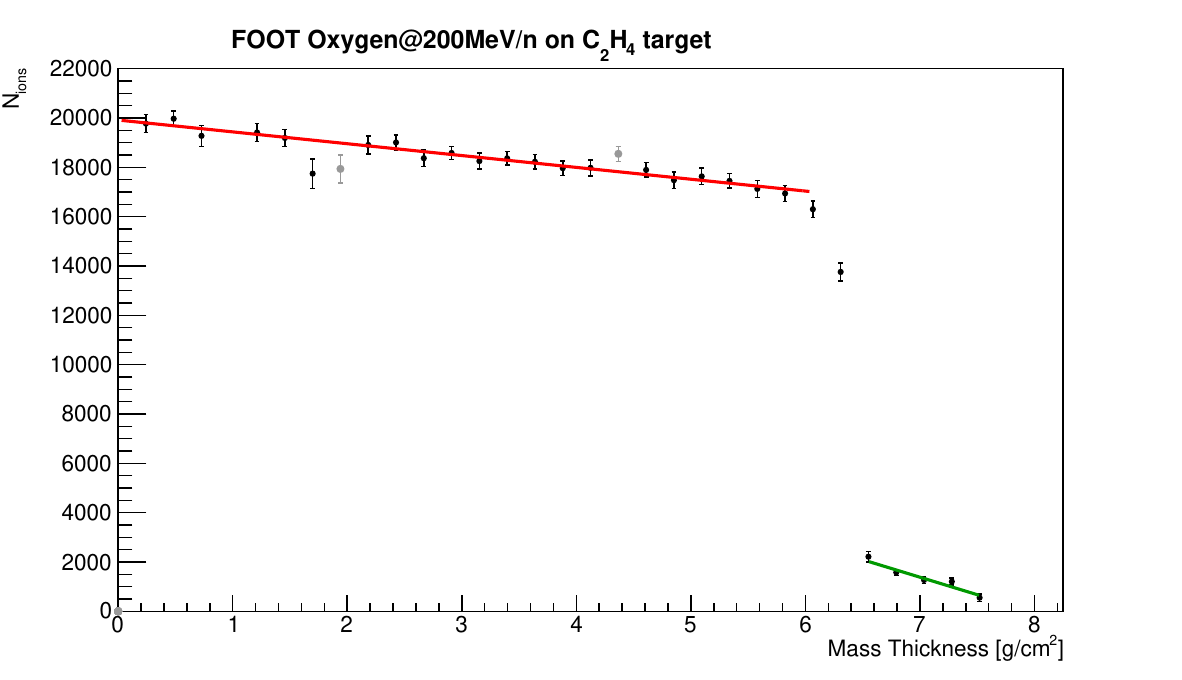}	
	\caption{Number of surviving Oxygen ions and heavy fragments as a function of the mass thickness in ECC1 (left) and ECC2 (right).} 
	\label{fig:oxy_fit}
\end{figure}

The total number of detected ions, which includes both primary Oxygen ions and heavy fragments, is shown in Figure~\ref{fig:oxy_fit} as a function of the cumulative mass thickness of the traversed emulsion material. 
The total number of detected ions can be well described by a linear dependence on the cumulative mass thickness of the traversed emulsion material, as in the MC simulation. The observed linear dependence indicates that the number of interactions remains constant with the traversed mass thickness, but since the number of primary ions gradually decreases along the stack, this behavior is consistent with a weak energy dependence of the total reaction cross-section, as also observed in MC simulations. Films with significantly lower efficiency, sporadically observed along the stack, identified and shaded in gray, were excluded from the fit.  

The evaluated number of Oxygen ions at the entrance of the ECC agrees with the measurement from the Start Counter detector within 1\%.

The contribution of heavy fragments must be accounted for in order to accurately estimate $N_{B}$. These fragments are produced by inelastic nuclear interactions and can traverse emulsion films beyond the Bragg peak of the primary beam. In the case of ECC1, the Bragg peak occurs at a cumulative mass thickness of about 7 g/cm$^2$, in the second section of the detector, which contains no passive material. As a result, the number of fragments remains approximately constant beyond this point, as shown in Figure~\ref{fig:oxy_fit} (top). In ECC2, in contrast, the Bragg peak is reached in the first section, at around 6.2 g/cm$^2$, as evident in Figure~\ref{fig:oxy_fit} (bottom). 

Assuming that the heavy fragments were produced uniformly across the first section of the detector, their average production rate per unit mass thickness can be expressed as $\frac{N_{\text{heavy}}}{T} $, where \(N_{\text{heavy}}\) denotes the total number of heavy fragments produced prior to the Bragg peak, and \(T\) represents the cumulative mass thickness traversed by the primary particles before stopping. \(N_{\text{heavy}}\) was measured from the emulsion films immediately after the Bragg peak. For ECC2, a linear fit (green line in figure~\ref{fig:oxy_fit}) was performed and it was extrapolated up to the last plate included in the fit containing the primary Oxygens (red line). For ECC1, the limited amount of passive material prevented a reliable extrapolation; therefore, the number of fragments in the first available untreated emulsion film after the Bragg peak was used. These contributions were subtracted from the slope $p_{1}$ indicated to isolate the contribution of the Oxygen ions, thus obtaining $p_{oxy}' = p_{1} - \frac{N_{heavy}}{T}$. The procedure was repeated in the MC sample. 
The MC validation of the method showed that the reconstructed slope $p_{oxy}'$ underestimated the true one by approximately 10\% in both ECC1 and ECC2. Therefore, the final estimate was obtained by subtracting the contribution of the heavy fragments and correcting for this effect: $p_{oxy} = \frac{p_{1} - N_{heavy}/T}{\eta_{oxy}}$, with $\eta_{oxy}$ equal to $90.7 \,\%\pm 0.2\,\%$ for ECC2 and  to $88.0 \,\% \pm 0.1 \,\%$ for ECC1. 

Based on this analysis, we estimate that 27\% and 33\% of primary Oxygen ions underwent nuclear interactions before stopping in ECC1 and ECC2, respectively. The equations of the linear fit describing the disappearance of the number of Oxygen ions are:

\begin{equation}
    N_B = (-800 \pm 60) \frac{\text{cm}^2}{\text{g}} \cdot \text{Mass Thickness} + (19530\pm190)
\end{equation}
\begin{equation}
    N_B = (-1100\pm 60) \frac{\text{cm}^2}{\text{g}} \cdot \text{Mass Thickness} + (20200\pm150)
\end{equation}

for ECC1 and ECC2 respectively. The extraction of the reaction cross-section from the slope of disappearing Oxygens is not possible, since from such an analysis we could not distinguish the ions interacting in the target layers from those interacting in the nuclear emulsion films, which would introduce a significant systematic uncertainty.

\subsection{$\epsilon_{reco}$ evaluation}\label{eff}

The reconstruction factor $\epsilon_{reco}$ was defined as the ratio between the number of a reconstructed quantity in the MC sample and the corresponding true value:

\begin{equation}
\epsilon_{reco} (x) = \frac{Y^{\text{MC Reconstructed}}(x)}{{Y^{\text{MC True}}}(x)}
\end{equation}
This factor provides an indication of the quality of the reconstruction. 


\subsection{Systematic error evaluation}\label{syst}

To evaluate potential systematic uncertainties, a study was conducted comparing the number of reconstructed interaction vertices ($Y_{V}$) with the number of primary Oxygen ions that stop in the material ($N_{B}$), both expressed as functions of the product of material thickness and density. In this analysis, interactions occurring within the nuclear emulsion films, which were excluded from the cross-section evaluation, are also taken into account.

\begin{figure}
	\includegraphics[width=0.54\textwidth]{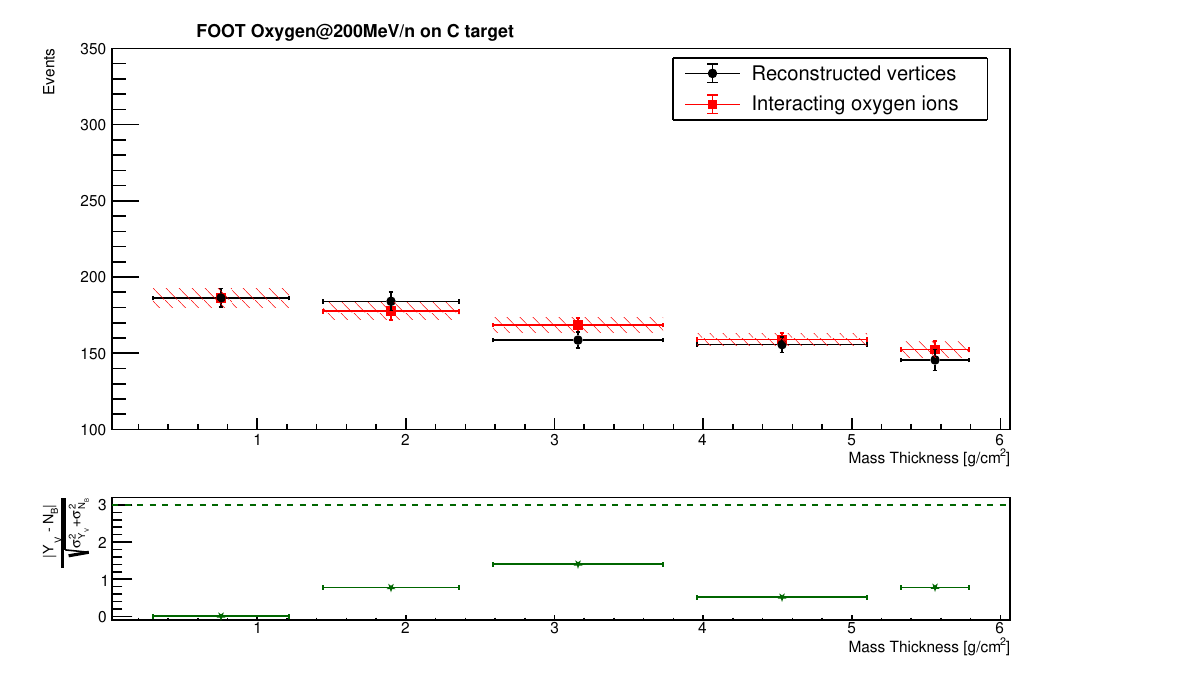}	
    	\includegraphics[width=0.54\textwidth]{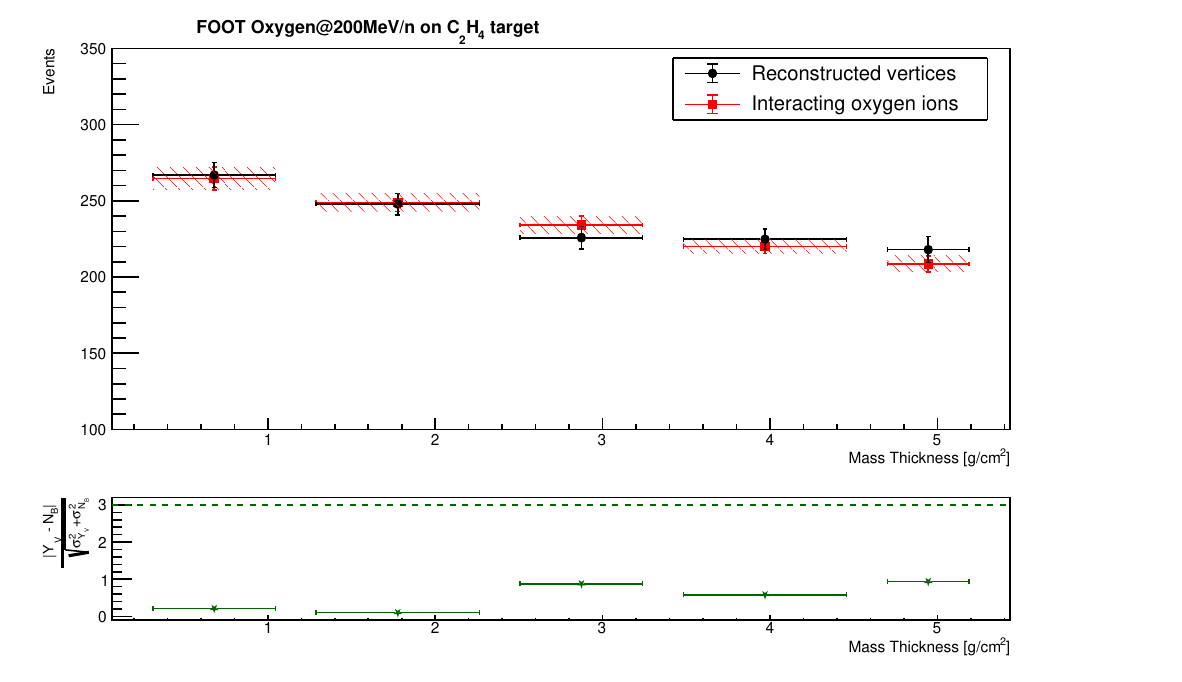}	
	\caption{Number of reconstructed vertices compared to the number of interacting Oxygen ions for C (left) and C$_{2}$H$_{4}$ (right) targets. The bottom sub-panel in each case presents the normalized difference $(|Y_{V} - N_{B}|)/\sqrt{\sigma_{Y_{V}}^2 + \sigma_{N_{B}}^2}$ as a measure of the discrepancy.} 
	\label{fig:oxy_disap_C}
\end{figure}

The methods used to determine these observables are entirely independent, both in terms of hardware and software, as described in Sections~\ref{Y} and~\ref{NB}, thus providing a robust basis for comparison. The results for ECC1 and ECC2 are shown in Figure~\ref{fig:oxy_disap_C}, with error bars representing one standard deviation. For each data point, the normalized difference between the two measurements, defined as $(|Y_{V} - N_{B}|)/\sqrt{\sigma_{Y_{V}}^2 + \sigma_{N_{B}}^2}$, consistently remains below~1.5, as shown in the lower sub-panels of Figure~\ref{fig:oxy_disap_C}. Furthermore, the comparison reveals no systematic behavior in the observed differences between the two estimates. 
This observation indicates that the systematic uncertainty is smaller than the statistical one and can thus be considered negligible for the purpose of the cross-section evaluation~\citep{ParticleDataGroup:2024}.

\section{Results and discussion}\label{sec2}

\begin{figure}
	\centering 
	\includegraphics[width=0.8\textwidth]{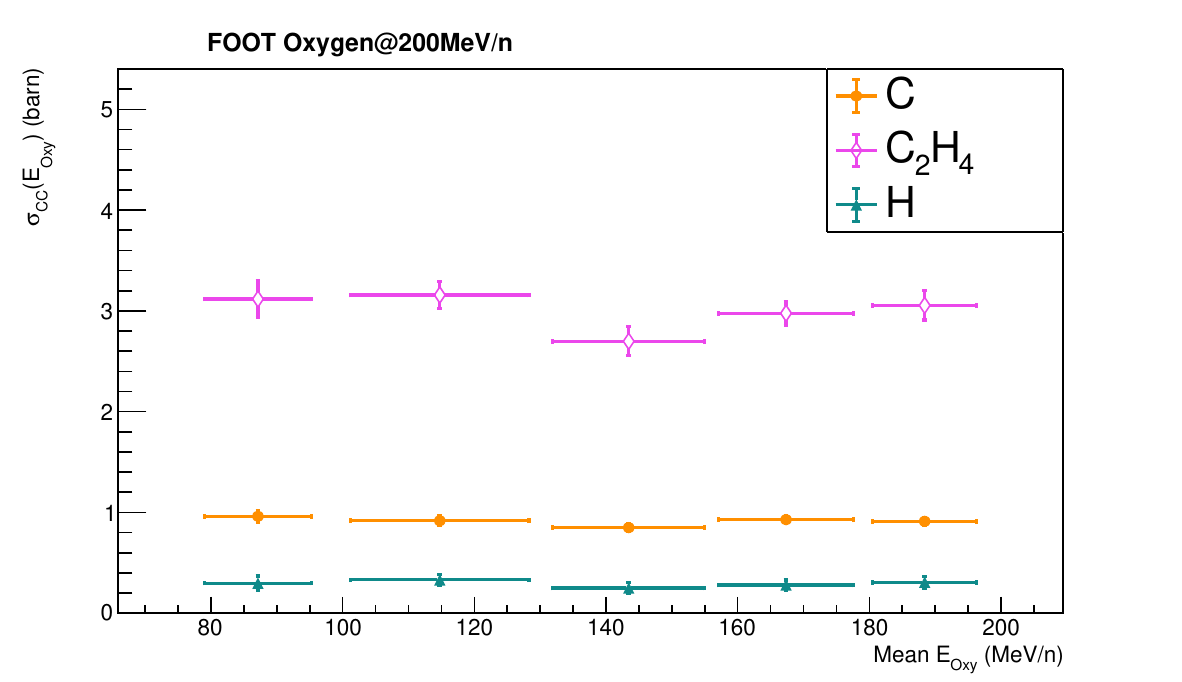}	
	\caption{Total charge-changing cross-sections on C, C$_2$H$_4$ and H as a function of energy.} 
	\label{fig:xsec_vtx}
\end{figure}

The total charge-changing cross-section $\sigma_{CC}$ as a function of energy is reported in Figure~\ref{fig:xsec_vtx}. The corresponding numerical values are detailed in Table~\ref{tab:xsec_vtx}. The beam energy in each module was estimated by the calculation of the energy loss using FLUKA MC simulation~\citep{osti_877507,Fluka2:2016}.

\begin{table}[h]
\caption{\label{tab:xsec_vtx} Measured total charge-changing cross-sections on C, C$_2$H$_4$ and H across different energy intervals.}
\begin{tabular}{@{}llll@{}}
\toprule
\textbf{Beam Ekin} & \textbf{\boldmath$\sigma_{CC}$ on C} & \textbf{\boldmath$\sigma_{CC}$ on C$_2$H$_4$} & \textbf{\boldmath$\sigma_{CC}$ on H} \\
        \textbf{(MeV/n)} & \textbf{(barn)} & \textbf{(barn)} & \textbf{(barn)} \\
\midrule
 $188\pm 7$ & ${0.91\pm 0.04}$  &  ${3.0\pm0.1}$ & ${0.31\pm0.06}$\\
 $167\pm10$ & ${0.93\pm 0.04}$ &  ${3.0\pm0.1}$ & ${0.28\pm0.05}$\\
$ 143\pm 11$ & ${0.85 \pm 0.04}$ &  ${2.7\pm0.1}$ & ${0.25\pm0.06}$\\
 $115\pm 14$ & ${0.92\pm0.05}$ &  ${3.2\pm0.1}$ & ${0.33\pm0.06}$\\
 $87\pm8$ & ${0.96\pm 0.06}$  &  ${3.1\pm0.2}$ & ${0.30\pm0.08}$\\
\botrule
\end{tabular}
\end{table}



\begin{figure}
	\centering 
	\includegraphics[width=0.8\textwidth]{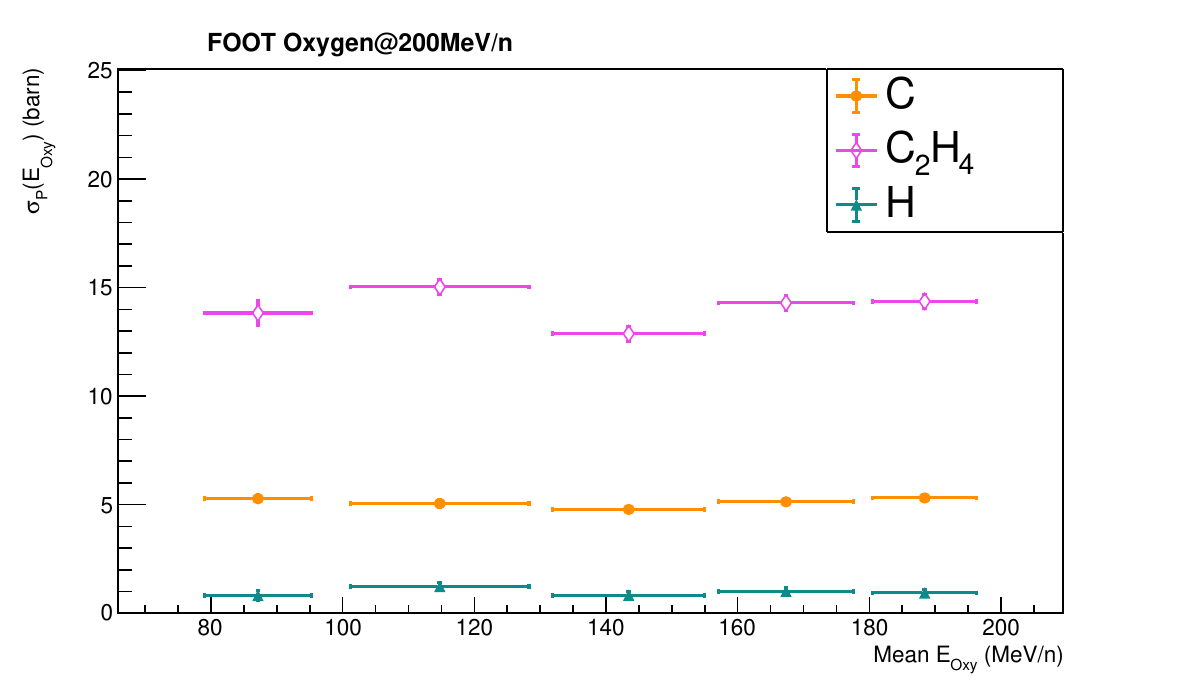}	
	\caption{Fragment production cross-sections on C, C$_2$H$_4$ and H as a function of energy.} 
	\label{fig:xsec_trk}
\end{figure}

The fragment production cross-section $\sigma_P$ as a function of energy is reported in Figure~\ref{fig:xsec_trk} for C, C$_2$H$_4$ and H targets. The corresponding numerical values are reported in Table~\ref{tab:xsec_trk}.

\begin{table}[h]
\caption{\label{tab:xsec_trk} Measured fragment production cross-sections on C, C$_2$H$_4$ and H across different energy intervals.}
\begin{tabular}{@{}llll@{}}
\toprule
\textbf{Beam Ekin} & \textbf{$\sigma_P$ on C} & \textbf{$\sigma_P$ on C$_2$H$_4$} & \textbf{$\sigma_P$ on H} \\
        \textbf{(MeV/n)} & \textbf{(barn)} & \textbf{(barn)} & \textbf{(barn)} \\
\midrule
 $188\pm 7$ & ${5.3\pm 0.1}$  &  ${14.4\pm0.3}$ & ${0.93\pm0.14}$\\
 $167\pm10$ & ${5.1\pm 0.2}$ &  ${14.3\pm0.4}$ & ${1.01\pm0.17}$\\
$ 143\pm 11$ & ${4.8 \pm 0.1}$ &  ${12.9\pm0.3}$ & ${0.83\pm0.18}$\\
 $115\pm 14$ & ${5.1\pm0.2}$ &  ${15.0\pm0.4}$ & ${1.23\pm0.18}$\\
 $87\pm8$ & ${5.3\pm 0.2}$  &  ${13.8\pm0.6}$ & ${0.82\pm0.23}$\\
 \botrule
\end{tabular}
\end{table}


Previous measurements of $^{16}$O fragment production cross-sections~\cite{Yamaguchi2011, Zeitlin2011, Webber1990} are summarized in Table~\ref{tab:comparison_C} for a C target and in Table~\ref{tab:comparison_C2H4} for a CH$_2$ target. Note that the cross-section on C$_2$H$_4$ is twice that on CH$_2$, due to the different molecular composition.

\begin{table}[h]
\caption{\label{tab:comparison_C} Available results from other experiments for $\sigma_{CC}$ on C.}
\begin{tabular}{@{}llll@{}}
\toprule
 \textbf{Ref.} & \textbf{Beam Ekin} & \textbf{Angular} & \textbf{\boldmath$\sigma_{CC}$ on C}\\
  &  \textbf{(MeV/n)} & \textbf{Acceptance} & \textbf{(barn)}\\
\midrule
        Yamaguchi~\cite{Yamaguchi2011} & 288  & 10$^{\circ}$ & 0.852$\pm$0.017\\ 
        Zeitlin~\cite{Zeitlin2011} & 290  & 5.7$^{\circ}$ & 0.863$\pm$0.020\\ 
        Zeitlin~\cite{Zeitlin2011} & 400 & 6.7$^{\circ}$ & 0.842$\pm$0.022 \\ 
\botrule
\end{tabular}
\end{table}

\begin{table}[h]
\caption{\label{tab:comparison_C2H4} Available results from other experiments for $\sigma_{CC}$ on CH$_2$.}
\begin{tabular}{@{}llll@{}}
\toprule
  \textbf{Ref.} & \textbf{Beam Ekin} & \textbf{Angular} & \textbf{\boldmath$\sigma_{CC}$ on CH$_2$}\\
  &  \textbf{(MeV/n)} & \textbf{Acceptance} & \textbf{(barn)}\\
\midrule
    Webber~\cite{Webber1990} & 441 & 7.7$^{\circ}$ & 1.260$\pm$0.013 \\ 
    Webber~\cite{Webber1990} & 591 & 7.7$^{\circ}$ & 1.316$\pm$0.013 \\ 
    Webber~\cite{Webber1990} & 669 & 7.7$^{\circ}$ & 1.328$\pm$0.013 \\ 
\botrule
\end{tabular}
\end{table}

It is important to highlight that previous measurements were conducted at higher energies, with data on Carbon targets collected at energies at least 100 MeV/n above those covered in this study, and those on CH$_2$ targets exceeding our energy range by more than 250 MeV/n. Furthermore, angular acceptance was more limited, typically below 10$^{\circ}$, as detailed in Tables~\ref{tab:comparison_C} and~\ref{tab:comparison_C2H4}. In the present work, fragments are detected up to polar angles corresponding to $\tan\theta = 1$ (i.e. 45$^{\circ}$), thus complementing existing data by covering a wider angular range and extending measurements to lower beam energies. 


\subsection{Comparison with prominent nuclear interaction models}\label{comparison}

\begin{figure}
	\centering 
	\includegraphics[width=0.8\textwidth]{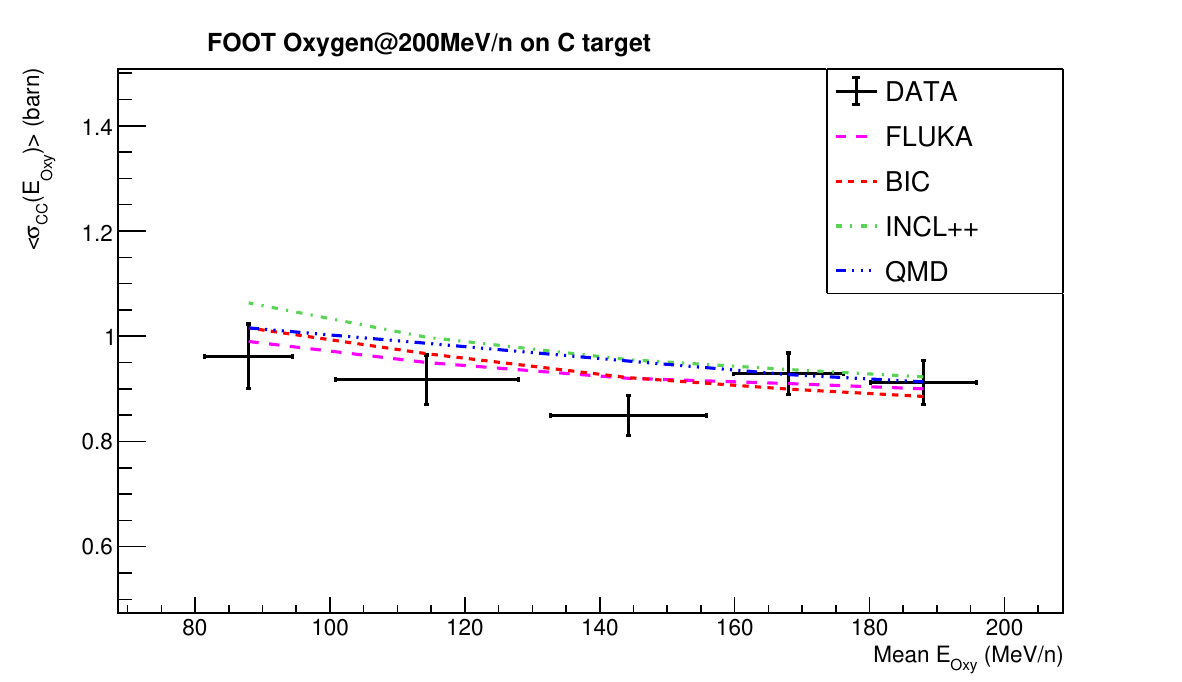}	
	\caption{Experimental total charge-changing cross-sections on C as a function of energy together with FLUKA and Geant4 predictions with four different models.} 
	\label{fig:xsec_geant_C}
\end{figure}

\begin{figure}
	\centering 
	\includegraphics[width=0.8\textwidth]{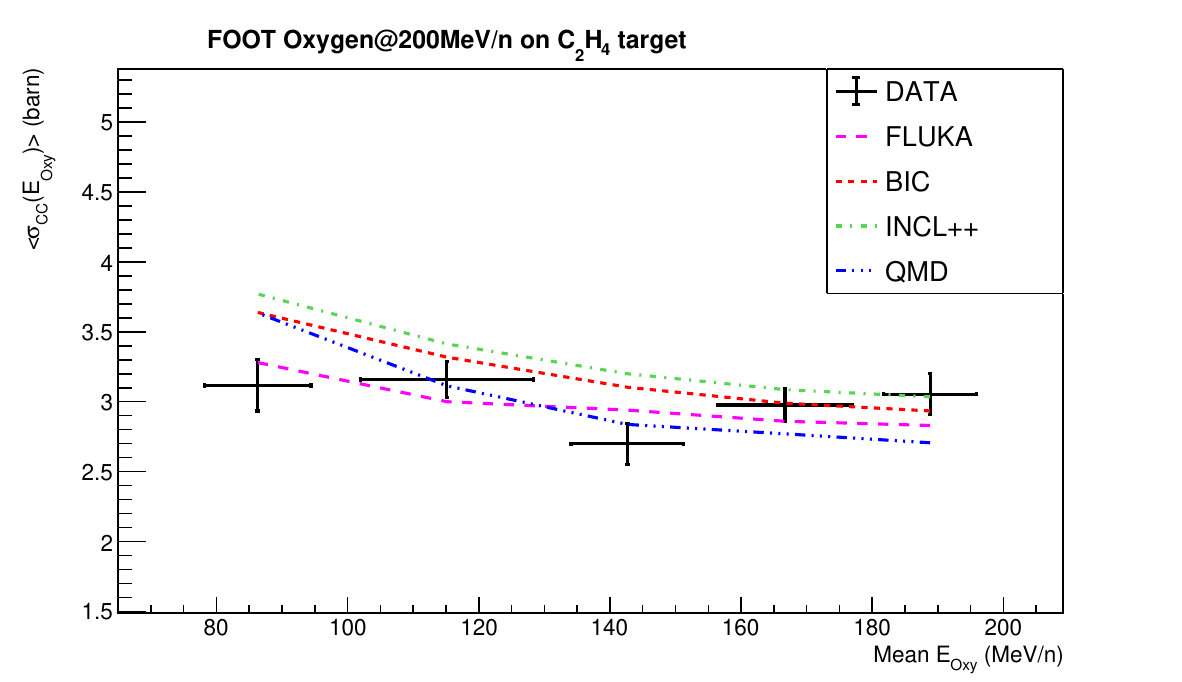}	
	\caption{Experimental total charge-changing cross-sections on C$_2$H$_4$ as a function of energy together with FLUKA and Geant4 predictions with four different models.} 
	\label{fig:xsec_geant_C2H4}
\end{figure}

The experimental charge-changing cross-sections were compared with theoretical predictions of four different nuclear interaction models used in MC codes, as shown in Fig.~\ref{fig:xsec_geant_C} and~\ref{fig:xsec_geant_C2H4} for C and C$_2$H$_4$ respectively. The models considered include FLUKA~\cite{Fluka2:2016} and three different hadronic models implemented within the Geant4 framework~\cite{AGOSTINELLI2003250,geant4_2,ALLISON2016186} hadronic models: Binary Ion Cascade (BIC)~\cite{Wellisch:2005zz}, Quantum Molecular Dynamics (QMD)~\cite{PhysRevC.79.014614}, and Li\`ege Intranuclear Cascade (INCL++)~\cite{PhysRevC.90.054602}.

The BIC model was built through the \texttt{G4IonPhysics} physics constructor, specifically designed for ion-ion collisions. It combines several models to simulate interactions across a broad energy range. For light and heavy ions (Z $>$ 2), it uses Glauber-Gribov parameterizations for cross-sections and couples the BIC model with a Precompound model for low-energy de-excitation. At higher energies (above 3~GeV/nucleon), the interaction is handled by the FTFP model, which incorporates string excitation followed by precompound de-excitation.

QMD was accessed via the \texttt{G4IonQMDPhysics} library, frequently used in applications involving light-ion transport, such as medical physics and space radiation studies. It provides a microscopic approach in which nucleons are treated as Gaussian wave packets, whose evolution follows the quantum molecular dynamics framework. This allows for a more detailed treatment of the dynamical evolution of nuclear collisions compared to purely phenomenological models.

Both BIC and QMD are used in conjunction with the \texttt{G4HadronPhysicsQGSP\_BIC} physics list, which integrates the Quark-Gluon String Precompound (QGSP) model for high-energy hadronic interactions with the BIC cascade for the low-energy regime. This combined configuration is widely used in various simulation contexts, including high-energy and medical applications.

The third Geant4 model considered is INCL++, built through the \texttt{G4IonINCLXXPhysics} physics list, which simulates nuclear reactions as a sequence of individual nucleon-nucleon interactions inside the target nucleus. It models the nucleus as a Fermi gas of nucleons contained in a static potential well, and distinguishes between projectile participants (entering the target volume) and spectators. The resulting fragmentation is treated asymmetrically: the quasi-projectile is formed from the projectile spectators and non-interacting participants, while the quasi-target evolves from the full cascade dynamics. This model is particularly effective at reproducing target fragmentation and is typically run in inverse kinematics to improve the modeling of projectile fragmentation. However, when both the projectile and the target have mass number A$ >$ 18, the simulation defaults to using the BIC model instead.

General agreement in the overall trend of the measured and simulated distributions is observed across all models. 

For the Carbon target, shown in Fig.~\ref{fig:xsec_geant_C}, the overall agreement between data and model predictions is reasonable across the explored energy range. FLUKA and BIC show the closest agreement in both shape and magnitude, while INCL++ remains higher over the full range. In particular, FLUKA reproduces both the shape and magnitude of the measured cross-sections with good accuracy, while BIC slightly overestimates the data below about 140~MeV/n but aligns well at higher energies. QMD, on the other hand, predicts a flatter energy dependence and tends to overestimate the cross-sections at the lowest energies. These comparisons indicate that, despite residual differences, all models capture the energy dependence of the measured cross-sections, with FLUKA providing the most accurate description.

In the case of the Polyethylene target, shown in Fig.~\ref{fig:xsec_geant_C2H4}, the experimental data are overall well reproduced by FLUKA across the full energy range, showing good agreement in both shape and normalization. QMD exhibits a similar behavior above 100~MeV/n, while slightly overestimating the data at lower energies. Both BIC and INCL++ tend to overestimate the measured cross-sections below about 160~MeV/n but provide a better description at higher energies, gradually converging toward the experimental values.

Some of the observed deviations may be influenced by statistical fluctuations. Despite the limited statistics, the present results already provide constraints for model comparison and validation. Future measurements with increased statistics will further reduce uncertainties and strengthen the quantitative assessment of the different nuclear interaction models. 


\section{Conclusion}
We have measured the total charge-changing and fragment production cross-sections for interactions of $^{16}$O ions with Carbon and Polyethylene targets in the 80--200~MeV/n energy range using the nuclear emulsion spectrometers of the FOOT experiment. Moreover, the total charge-changing and fragment production cross-sections for $^{16}$O on Hydrogen in the same energy range are derived. These results represent the first measurement in this energy range with a large angular acceptance, up to 45$^\circ$. Results were compared with previous experimental results and MC model predictions. Overall, the experimental data show good agreement with the model predictions, with FLUKA providing the closest description. 


These measurements provide new benchmark data in an energy region where experimental information remains scarce. They contribute to the validation of transport models employed in particle therapy and represent the first step toward differential fragmentation measurements with the FOOT emulsion spectrometer.



Future developments will focus on differential measurements as functions of fragment charge and kinetic energy, together with an increased statistical sample.  


\section{Acknowledgements}
This work was supported  within the frame of FAIR Phase-0, Bio-PAC 2020 (proposal SBio08\_Battistoni), the GSI Helmholzzentrum für Schwerionenforschung, and by the European Union - Next Generation EU (Mission 4 Component 1, CUP H53D23001090006). 
The FOOT Collaboration acknowledges the INFN for its support in building and running the detector.
We would like to acknowledge all the personnel of GSI centre who provided us with support during data collection.


\bibliography{biblio}

\end{document}

%% file: authors.tex
\author*[1,2]{\fnm{Giuliana}\sur{Galati}}
\email{giuliana.galati@uniba.it}

\author*[3]{\fnm{Vincenzo}\sur{Boccia}}
\email{vincenzo.boccia@na.infn.it}

\author[3]{\fnm{Andrey}\sur{Alexandrov}}
\author[4]{\fnm{Giovanni}\sur{Ambrosi}}
\author[5,6]{\fnm{Stefano}\sur{Argir\`o}}
\author[7]{\fnm{Takashi}~\sur{Asada}}
\author[4]{\fnm{Mattia}\sur{Barbanera}}
\author[8,6]{\fnm{Nazar}\sur{Bartosik}}
\author[9]{\fnm{Giuseppe}\sur{Battistoni}}
\author[10,11]{\fnm{Alexandre}\sur{Bigot}}
\author[12,13]{\fnm{Maria Giuseppina}\sur{Bisogni}}
\author[14]{\fnm{Giorgio}~\sur{Butella}}
\author[6]{\fnm{Francesca}~\sur{Cavanna}}
\author[6]{\fnm{Piergiorgio}~\sur{Cerello}}
\author[12,13]{\fnm{Esther}~\sur{Ciarrocchi}}
\author[15]{\fnm{Nicola}~\sur{D'Ambrosio}}
\author[3,16]{\fnm{Giovanni}~\sur{De Lellis}}
\author[3,16]{\fnm{Antonia}~\sur{Di Crescenzo}}
\author[17,2]{\fnm{Benedetto}~\sur{Di Ruzza}}
\equalcont{deceased}

\author[18,19]{\fnm{Matilde}~\sur{Dondi}}
\author[14]{\fnm{Marco}~\sur{Donetti}}
\author[9]{\fnm{Yunsheng}~\sur{Dong}}
\author[16,20,21]{\fnm{Marco}~\sur{Durante}}
\author[22,23]{\fnm{Riccardo}~\sur{Faccini}}
\author[6,5]{\fnm{Veronica}~\sur{Ferrero}}
\author[10]{\fnm{Christian}~\sur{Finck}}
\author[6]{\fnm{Elisa}~\sur{Fiorina}}
\author[3]{\fnm{Marco}~\sur{Francesconi}}
\author[19,18]{\fnm{Matteo}~\sur{Franchini}}
\author[24,23]{\fnm{Gaia}~\sur{Franciosini}}
\author[13]{\fnm{Luca}~\sur{Galli}}
\author[16]{\fnm{Antonio}~\sur{Iuliano}}
\author[4]{\fnm{Keida}~\sur{Kanxheri}}
\author[6]{\fnm{Bharat}~\sur{Kharpuse}}
\author[13]{\fnm{Aafke}~\sur{Kraan}}
\author[3,16]{\fnm{Adele}~\sur{Lauria}}
\author[25,6]{\fnm{Ernesto}~\sur{Lopez Torrez}}
\author[24,23]{\fnm{Marco}~\sur{Magi}}
\author[18]{\fnm{Alice}~\sur{Manna}}
\author[26,23]{\fnm{Michela}~\sur{Marafini}}
\author[15]{\fnm{Simone}~\sur{Masci}}
\author[13]{\fnm{Maurizio}~\sur{Massa}}
\author[18,19]{\fnm{Christian}~\sur{Massimi}}
\author[9]{\fnm{Ilaria}~\sur{Mattei}}
\author[4,27]{\fnm{Sofia}~\sur{Mazzolani}}
\author[18]{\fnm{Alberto}~\sur{Mengarelli}}
\author[14]{\fnm{Alessio}~\sur{Mereghetti}}
\author[22,23]{\fnm{Riccardo}~\sur{Mirabelli}}
\author[13]{\fnm{Andrea}~\sur{Moggi}}
\author[28,29]{\fnm{Maria Cristina}~\sur{Morone}}
\author[12,13]{\fnm{Matteo}~\sur{Morrocchi}}
\author[9]{\fnm{Silvia}~\sur{Muraro}}
\author[7,30]{\fnm{Tatsuhiro}~\sur{Naka}}
\author[6]{\fnm{Nadia}~\sur{Pastrone}}
\author[24,23]{\fnm{Vincenzo}~\sur{Patera}}
\author[6]{\fnm{Francesco}~\sur{Pennazio}}
\author[18,19]{\fnm{Claudia}~\sur{Pisanti}}
\author[4]{\fnm{Pisana}~\sur{Placidi}}
\author[14]{\fnm{Marco}~\sur{Pullia}}
\author[23,24]{\fnm{Flaminia}~\sur{Quattrini}}
\author[18]{\fnm{Sara}~\sur{Rabaglia}}
\author[31,6]{\fnm{Luciano}~\sur{Ramello}}
\author[20]{\fnm{Claire-Anne}~\sur{Reidel}}
\author[18]{\fnm{Riccardo}~\sur{Ridolfi}}
\author[28,29,26]{\fnm{Laura}~\sur{Rocchetti}}
\author[32]{\fnm{Lucia}~\sur{Sabbatini}}
\author[4]{\fnm{Lucia}~\sur{Salvi}}
\author[32]{\fnm{Claudio}~\sur{Sanelli}}
\author[24,23]{\fnm{Alessio}~\sur{Sarti}}
\author[30]{\fnm{Osamu}~\sur{Sato}}
\author[14]{\fnm{Simone}~\sur{Savazzi}}
\author[24,23]{\fnm{Angelo}~\sur{Schiavi}}
\author[20]{\fnm{Christoph}~\sur{Schuy}}
\author[33]{\fnm{Emanuele}~\sur{Scifoni}}
\author[4]{\fnm{Leonello}~\sur{Servoli}}
\author[4]{\fnm{Gianluigi}~\sur{Silvestre}}
\author[8,6]{\fnm{Mario}~\sur{Sitta}}
\author[5,6] {\fnm{Benedetto}~\sur{Spadavecchia}}
\author[18,19]{\fnm{Roberto}~\sur{Spighi}}
\author[32]{\fnm{Eleuterio}~\sur{Spiriti}}
\author[22,23,26]{\fnm{Luana}~\sur{Testa}}
\author[3]{\fnm{Valeri}~\sur{Tioukov}}
\author[32]{\fnm{Sandro}~\sur{Tomassini}}
\author[34,33]{\fnm{Francesco}~\sur{Tommasino}}
\author[24,23]{\fnm{Marco}~\sur{Toppi}}
\author[23]{\fnm{Giacomo}~\sur{Traini}}
\author[32]{\fnm{Antonio}~\sur{Trigilio}}
\author[18,19]{\fnm{Giacomo}~\sur{Ubaldi}}
\author[18,19]{\fnm{Sara}~\sur{Valentinetti}}
\author[10]{\fnm{Marie}~\sur{Vanstalle}}
\author[18,19]{\fnm{Mauro}~\sur{Villa}}
\author[20]{\fnm{Uli}~\sur{Weber}}
\author[18,19]{\fnm{Roberto}~\sur{Zarrella}}
\author[18,19]{\fnm{Antonio}~\sur{Zoccoli}}
\author[35,3]{and \fnm{Maria Cristina}~\sur{Montesi}}

\affil[1]{\orgdiv{University of Bari, Department of Physics, Bari, Italy}}
\affil[2]{\orgdiv{INFN, Section of Bari, Bari, Italy}}
\affil[3]{\orgdiv{INFN, Section of Napoli, Napoli, Italy}}
\affil[4]{\orgdiv{INFN, Section of Perugia, Perugia, Italy}}
\affil[5]{\orgdiv{University of Torino, Department of Physics, Torino, Italy}}
\affil[6]{\orgdiv{INFN, Section of Torino, Torino, Italy}}
\affil[7]{\orgdiv{Department of Physics, Faculty of Science, Toho University, Chiba, Japan}}

\affil[8]{\orgdiv{University of Piemonte Orientale, Department of Science and Technological Innovation, Alessandria, Italy}}
\affil[9]{\orgdiv{INFN, Section of Milano, Torino, Milano}}
\affil[10]{\orgdiv{Universit\'e de Strasbourg, CNRS, IPHC UMR 7871, F-67000 Strasbourg, France}}
\affil[11]{\orgdiv{IPHC, France}}
\affil[12]{\orgdiv{University of Pisa, Department of Physics, Pisa, Italy}}
\affil[13]{\orgdiv{INFN Section of Pisa, Pisa, Italy}}
\affil[14]{\orgdiv{CNAO, National Center for Oncological Hadrontherapy, Pavia, Italy}}
\affil[15]{\orgdiv{INFN, Section of Gran Sasso National Laboratories, Assergi (L’Aquila), Italy}}
\affil[16]{\orgdiv{University of Napoli, Department of Physics  "E.~Pancini", Napoli, Italy}}
\affil[17]{\orgdiv{University of Foggia, Foggia, Italy}}
\affil[18]{\orgdiv{INFN Section of Bologna, Bologna, Italy}}
\affil[19]{\orgdiv{University of Bologna, Department of Physics and Astronomy, Bologna, Italy}}
\affil[20]{\orgdiv{Biophysics Department, GSI Helmholtzzentrum f\"ur Schwerionenforschung, Darmstadt, Germany}}
\affil[21]{\orgdiv{Institute of Condensed Matter Physics, Technische Universit\"at Darmstadt, Darmstadt, Germany}}
\affil[22]{\orgdiv{Sapienza University of Rome, Department of Physics, Rome,Italy}}
\affil[23]{\orgdiv{INFN Section of Rome,  Rome, Italy}}
\affil[24]{\orgdiv{Sapienza University of Rome, Department of Basic and Applied Sciences for Engineering, Rome, Italy}}
\affil[25]{\orgdiv{CEADEN, Centro de Aplicaciones Tecnologicas y Desarrollo Nuclear, Havana, Cuba}}
\affil[26]{\orgdiv{Museo Storico della Fisica e Centro Studi e Ricerche Enrico Fermi, Rome, Italy}}
\affil[27]{\orgdiv{University of Camerino, Department of Physics, Camerino, Italy}}
\affil[28]{\orgdiv{University of Roma Tor Vergata, Physics Department, Rome, Italy}}
\affil[29]{\orgdiv{INFN Section of Roma Tor Vergata, Rome, Italy}}
\affil[30]{\orgdiv{Kobayashi-Maskawa Institute, Nagoya University, Nagoya, Japan}}
\affil[31]{\orgdiv{University of Piemonte Orientale, Department for Sustainable Development and Ecological Transition, Vercelli, Italy }}
\affil[32]{\orgdiv{INFN, Laboratori Nazionali di Frascati, Frascati, Italy}}
\affil[33]{\orgdiv{Trento Institute for Fundamental Physics and Applications, Istituto Nazionale di Fisica Nucleare (TIFPA-INFN), Trento, Italy}}
\affil[34]{\orgdiv{University of Trento, Physics Department, Trento, Italy}}
\affil[35]{\orgdiv{University of Napoli, Department of Chemistry, Napoli, Italy}}